\documentclass[twocolumn]{aastex63}

\usepackage{graphicx}

\usepackage{xspace}
\newcommand{\hMpc}{$h^{-1}$Mpc\xspace}
\newcommand{\logOH}{$12 + \log(\text{O}/\text{H})$\xspace}
\newcommand{\Mratio}{$M_{\text{DM}}/M_*$\xspace}

\newcommand{\Nvoid}{642\xspace}
\newcommand{\Nwall}{938\xspace}
\newcommand{\Nedge}{846\xspace}

\shorttitle{Mass ratio of void galaxies}
\shortauthors{Douglass, Smith, and Demina}

\begin{document}

\title{The influence of the void environment on the ratio of dark matter halo mass to stellar mass in SDSS MaNGA galaxies}

\correspondingauthor{Kelly A. Douglass}
\email{kellyadouglass@rochester.edu}

\author[0000-0002-9540-546X]{Kelly A. Douglass}
\affiliation{Department of Physics \& Astronomy, University of Rochester, 500 Wilson Blvd., Rochester, NY  14611, USA}

\author[0000-0003-3338-1105]{Jacob A. Smith}
\affiliation{Department of Physics \& Astronomy, University of Rochester, 500 Wilson Blvd., Rochester, NY  14611, USA}

\author[0000-0002-7852-167X]{Regina Demina}
\affiliation{Department of Physics \& Astronomy, University of Rochester, 500 Wilson Blvd., Rochester, NY  14611, USA}


\begin{abstract}
  We study how the void environment affects the formation and evolution of 
  galaxies in the universe by comparing the ratio of dark matter halo mass to 
  stellar mass of galaxies in voids with galaxies in denser regions.  Using 
  spectroscopic observations from the Sloan Digital Sky Survey MaNGA DR15, we 
  estimate the dark matter halo mass of \Nvoid void galaxies and \Nwall galaxies 
  in denser regions.  We use the relative velocities of the H$\alpha$ emission 
  line across the galaxy's surface to measure the rotation curve of each galaxy 
  because the kinematics of the interstellar medium is smoother than the stellar 
  kinematics.  We find that neither the stellar-to-halo-mass relation nor the 
  relationship between the gas-phase metallicity and the ratio of dark matter 
  halo mass to stellar mass is affected by the void environment.  We also 
  observe no difference in the distribution of the ratio of dark matter halo 
  mass to stellar mass between void galaxies and galaxies in denser regions, 
  implying that the shape of the dark matter halo profile is independent of a 
  galaxy's environment.
\end{abstract}


\section{Introduction}

As shown by large galaxy redshift surveys, the large-scale structure of the 
universe is well described as a three-dimensional cosmic web \citep{Bond96}: 
thin filaments of galaxies connect galaxy clusters and surround voids.  These 
voids are large, under dense regions of space that occupy close to 60\% of the 
volume of the universe \citep{daCosta88,Geller89,Pan12}.  Over the last couple 
of decades, the Sloan Digital Sky Survey \citep[SDSS;][]{York00} has provided an 
unprecedented data set that has made possible the study of the influence of the 
large-scale environment on the formation and evolution of galaxies.

Cosmic voids are an important environment for studying galaxy formation 
\citep[see][for a review]{vandeWeygaert11} because the gravitational clustering 
within them proceeds as if in a very low-density universe.  According to the 
$\Lambda$CDM cosmology, galaxies formed in voids should have lower masses and be 
retarded in their star formation when compared to those in more dense 
environments \citep{Gottlober03,Goldberg05,Cen11}.

Previous studies of the properties of void galaxies have found that their 
characteristics differ from galaxies that reside in denser environments.  In 
general, void galaxies have lower stellar mass 
\citep{Croton05,Hoyle05,Moorman15}, are bluer and of a later type 
\citep{Grogin00,Rojas04,Patiri06,Park07,vonBendaBeckmann08,Hoyle12}, have higher 
specific star formation rates 
\citep{Rojas05,vonBendaBeckmann08,Moorman15,Beygu16}, 
and are more gas-rich \citep{Kreckel12} than galaxies in denser regions.

The ratio of dark matter halo mass to stellar mass in a galaxy should affect the 
properties that depend on a galaxy's gravitational potential well.  Deeper 
potential wells prevent a higher fraction of gas from escaping the galaxy due to 
supernova-driven winds, tidal stripping, etc.  The gas-phase metallicity of a 
galaxy (defined as the relative abundance of oxygen relative to hydrogen, 
\logOH) is a measure of the integrated star formation history of a galaxy.  
Galaxies typically follow the mass-metallicity relation \citep{Tremonti04}, 
where more massive galaxies have higher metallicities.  While a galaxy's 
metallicity should depend on its stellar mass, it should also increase with its 
total mass due to the corresponding deeper potential well.

Void galaxies are expected to have lower gas-phase metallicities than those in 
denser regions if they only recently started forming stars or have recently 
accreted unprocessed gas.  However, 
\cite{Mouhcine07,Cooper08,Nicholls14,Kreckel15,Douglass17a,Douglass17b} and 
\cite{Douglass18} found that void dwarf galaxies do not have systematically 
lower gas-phase metallicities than dwarf galaxies in denser regions.  
\cite{Douglass17b} and \cite{Douglass18} posit that this might be the result of 
larger ratios of dark matter halo mass to stellar mass in void galaxies, a trend 
observed in \cite{Tojeiro17} and predicted by 
\cite{Arkhipova07,Cen11,Jung14,Tonnesen15,Martizzi19}.  We aim to test this 
hypothesis by estimating the mass ratios for void galaxies and comparing them to 
the mass ratios of galaxies in denser regions.

The primary objective of this paper is to compare the ratios of dark matter halo 
mass to stellar mass in spiral galaxies to discern if the void environment has 
an influence on their mass composition.  In this paper, we utilize the SDSS Data 
Release 15 \citep[DR15;][]{SDSS15}, which includes the most recent data from 
the SDSS MaNGA integral field spectroscopic survey \citep{MaNGA}.

\section{SDSS MaNGA Data and Galaxy Selection}

Unlike previous SDSS surveys, the SDSS~MaNGA survey measures spectra across the 
face of each observed galaxy.  Placing a bundle of spectroscopic fibers (an 
Integral Field Unit, IFU) that vary in size from 12'' (19 fibers) to 32'' (127 
fibers) on each galaxy \citep{Drory15}, MaNGA will collect spectra of 10,000 
nearby galaxies in the northern sky by its conclusion.  The IFUs are fed to two 
dual-channel spectrographs that simultaneously cover a wavelength range of 
3600--10300\AA\ with a resolution of $\lambda/\Delta \lambda \sim 2000$.  We 
use the H$\alpha$ velocity map, $V$-band image, and stellar mass density map as 
processed by Pipe3D \citep{Sanchez16,Sanchez18}, an analysis pipeline designed 
to investigate the properties of the stars and ionized gas in integral field 
spectroscopic data.  Galactic inclination angles, axis ratios, and absolute 
magnitudes are taken from the NASA-Sloan Atlas \citep{Blanton11}.

\subsection{IFU spaxel mask}\label{sec:masking}

In order to filter out invalid spaxels, we apply a series of restrictions to the 
various data fields used.   We mask those spaxels with an error in the H$\alpha$ 
velocity of either zero or not a number, a flux and/or error in the $V$-band of 
zero, and an estimate of zero for the stellar mass density.  These requirements 
result in an average of 21\% of spaxels masked in each galaxy.  Of the 4815 
galaxies processed by Pipe3D, 333 (6.9\%) are completely masked.

We also remove those galaxies with noisy velocity maps by requiring galaxies to 
have a ``smoothness'' score less than 2.27.  This score is determined by summing 
the cosine distance ($1 - \cos \theta$) across all unmasked spaxels and their 
four adjacent unmasked neighbors, where $\theta$ is the angle between the 
spaxel's gradient and its neighbor's gradient.  This sum is then normalized by 
the total number of unmasked spaxels in the velocity map.  As described in 
Section \ref{sec:curve_cuts}, additional quality selection criteria are applied 
after fitting the rotation curves.

\subsection{Environment classification}

\begin{figure*}
    \centering
    \includegraphics[width=\textwidth]{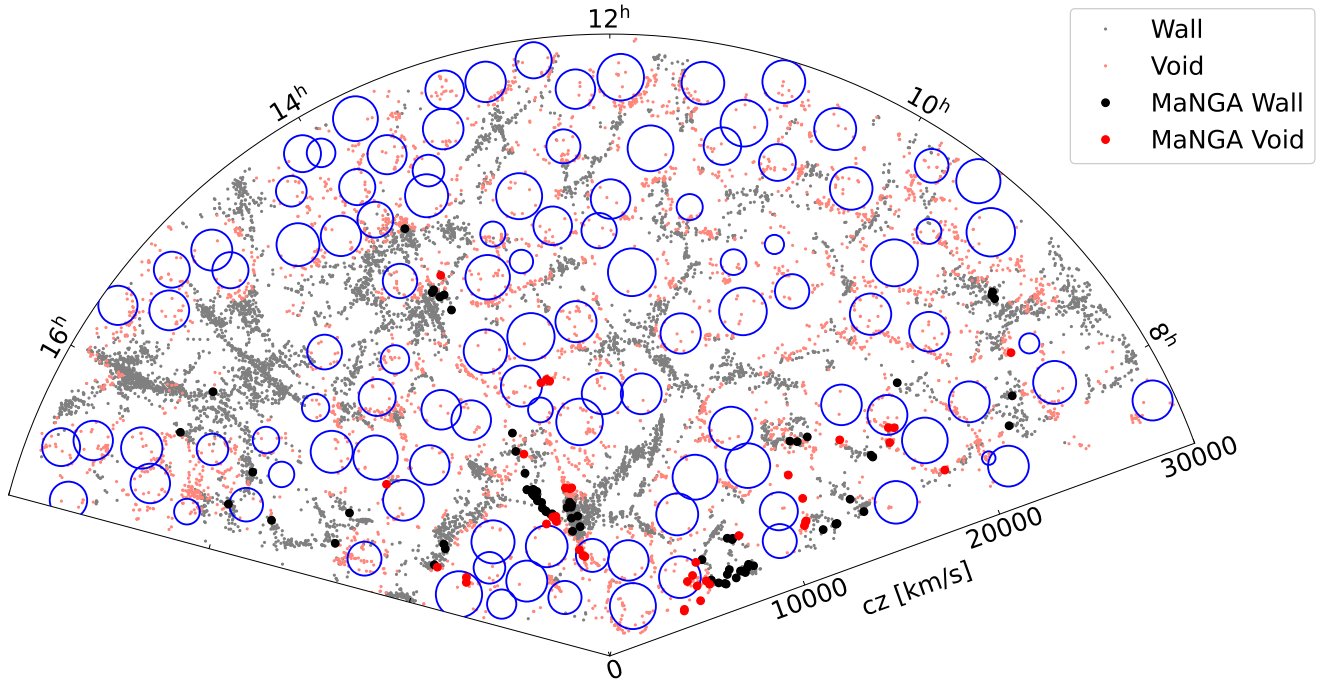}
    \caption{A $2^\circ$ thick decl. slice of the SDSS DR7 main galaxy sample.  
    The intersections of the maximal spheres of the voids with the center of 
    this slice are shown as blue circles.  The red points are galaxies that fall 
    within a void, while wall galaxies are shown in black.  The MaNGA galaxies 
    contained within this slice that are included in this study are identified 
    with larger points.  Note that as a result of only showing the maximal 
    spheres that intersect the midplane of the slice for the voids, some void 
    galaxies appear to fall outside the depicted voids.}
    \label{fig:wedge}
\end{figure*}

The large-scale environment for each galaxy is defined by the voids identified 
with VoidFinder \citep{Hoyle02}, an implementation of the void-finding algorithm 
described in \cite{ElAd97}, applied to a volume-limited sample of the SDSS~DR7 
galaxy catalog \citep{SDSS7} with absolute magnitude $M_r < -20$.  These voids 
are similar to the void catalog by \cite{Pan12} with more careful treatment of 
the survey edges.  VoidFinder removes all isolated galaxies, defined as those 
separated from their third nearest neighbor by more than 7\hMpc.  A grid is then 
applied to the remaining galaxies, and a sphere is grown from the center of each 
empty cell until it is bounded by four galaxies.  Each sphere with a minimum 
radius of 10\hMpc can serve as a seed for a void.   The sphere that seeds a void 
is referred to as that void's maximal sphere.  To identify dynamically distinct 
voids, maximal spheres are not permitted to overlap each other by more than 10\% 
of their volume.  If any of the remaining spheres overlap a maximal sphere by at 
least 50\% of its volume, it adds to that maximal sphere's void; otherwise it is 
discarded.  See \cite{Hoyle02} for a more detailed description of VoidFinder.

Galaxies that reside within a void are considered void galaxies; otherwise they 
are referred to as wall galaxies.  Due to the finite survey footprint of 
SDSS~DR7, we cannot locate voids within 10\hMpc of the survey boundary.  Any 
galaxy that falls within this border region or outside the survey footprint is 
classified as ``Uncertain.''  In Figure \ref{fig:wedge}, we show a $2^\circ$ 
thick decl. slice of the sky.  The maximal spheres that intersect the center of 
this slice are shown as the blue circles; galaxies that fall within a void are 
colored red, while those which make up the walls are in black.  The MaNGA 
galaxies that are part of our analysis are identified with larger points.

\section{Estimating the mass components}

The rotation curve is a measure of the galaxy's spin rate as a function of 
distance from the galaxy's center.  By measuring a galaxy's rotation curve, we 
can estimate the total mass of the galaxy.  Assuming Newtonian mechanics, the 
rotation rate of the stars and gas a distance $r$ from the center of the galaxy 
should only depend on the total enclosed mass within that distance, $M(r)$.  The 
following is an overview of the theory and method that we employ to estimate the 
galactic rotation curves in the MaNGA galaxies.

\subsection{Newtonian mechanics}

We assume that a galaxy's rotational motion is dominated by basic orbital 
mechanics: the orbital velocity of a particle some distance $r$ from the center 
of the galaxy is a function of the total mass internal to that radius, $M(r)$, 
assuming spherical symmetry.  For spiral galaxies, the orbital motion is assumed 
to be circular.  The gravitational force is the source of the centripetal 
acceleration for a particle in orbit, so
\begin{equation}\label{eqn:M_within_r}
    M(r) = \frac{v(r)^2 r}{G}
\end{equation}
Here, $v(r)$ is the velocity a distance $r$ from the center of the galaxy, and 
$G = 6.67408\times 10^{-11}$~m$^3$ kg$^{-1}$ s$^{-2}$ is the Newtonian 
gravitational constant.  Thus, by measuring $v(r)$ and $r$, we can estimate 
$M(r)$.

\subsection{Separating the mass components}

It is well known that the observed galaxy rotation curves for spiral galaxies do 
not match what we would expect based on the visible matter distribution 
\citep{Freeman70,Rubin70,Faber79,Rubin80,Bosma81b}.  If a galaxy's mass was 
composed entirely of baryonic matter, then its rotation curve would fall off as 
$1/\sqrt{r}$ beccause photometric analysis indicates that the majority of a 
galaxy's baryonic mass is located within the central part of the galaxy.  
However, the observed galaxy rotation curves level out, indicating that there is 
an additional component of material \citep{Freeman70,Rubin70}.  This additional, 
unobserved material has been dubbed ``dark matter,'' because it does not seem to 
emit light.

The rotation curve we measure from the data is a function of the total mass 
present within a given radius.  We assume that the two main components of the 
mass composition are dark matter and stellar mass.  Since the interstellar 
gas comprises only $\sim 10$\% of the disk's mass \citep{Nakanishi06}, including 
its contribution to the rotation curve will decrease our calculated dark matter 
mass by a negligible amount \citep{Paolo18}. The difference between the total 
and stellar mass, evaluated using an independent estimate (see Section 
\ref{sec:mass_decomp}), is our estimate for the dark matter mass at a given 
radius.

\section{Rotation Curve Analysis and Results}

\subsection{Measuring galactic rotation curves}

\begin{figure}
    \includegraphics[width=0.5\textwidth]{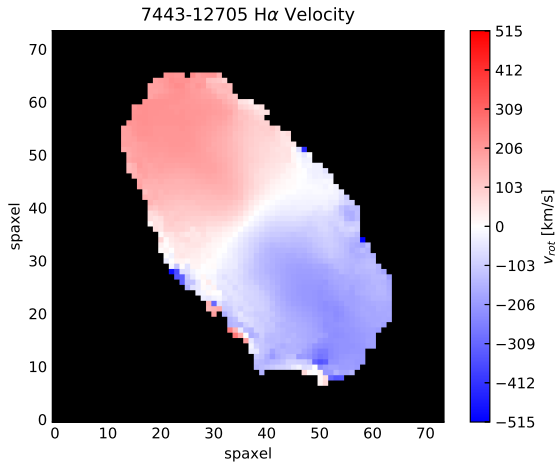}
    \caption{An example of the H$\alpha$ velocity field corrected to the 
    galaxy's rest frame.  Blue spaxels represent a blue-shifted H$\alpha$ 
    velocity, while red spaxels represent a red-shifted H$\alpha$ velocity.  
    The mask described in Section \ref{sec:masking} is applied.}
    \label{fig:Ha_vel_example}
\end{figure}

We use the velocity of the gas in the galaxy to determine its rotation curve, as 
gas kinematics are more continuous and less subject to random motions than the 
stellar population.  An example of the H$\alpha$ velocity map masked as 
described in Section \ref{sec:masking} is shown in Figure 
\ref{fig:Ha_vel_example}.  The velocities have been corrected to the internal 
motion of the galaxy by subtracting out the galaxy's center's velocity.  
It is unnecessary to recover the velocity perpendicular to our line of sight due 
to the assumption that the galaxy's stars and gas all rotate with the same 
tangential velocity at a given radius.  As a result, the maximum line-of-sight 
speed measured at a given radius of the galaxy corresponds only to the product 
of the tangential velocity at that radius and the sine of the inclination angle.  
The velocities have also been deprojected by assuming that the photometric 
inclination is the same as the kinetic inclination across the entire galaxy.  
As done in \cite{BarreraBallesteros18}, we assume that the kinetic and 
photometric centers of the galaxies are identical.

\begin{figure}
    \includegraphics[width=0.5\textwidth]{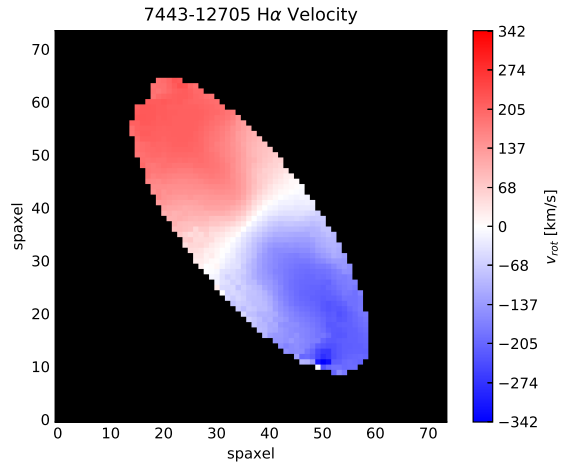}
    \caption{An example of the H$\alpha$ velocity spaxels within all the annuli 
    that are used in the analysis.  Due to the galaxy's angle of inclination, 
    the annuli are not necessarily circular from our perspective.}
    \label{fig:contour_plot_example}
\end{figure}

By taking into account the galaxy's angle of inclination and axis ratio, we find 
the minimum and maximum velocities along a given annulus of the galaxy.  This 
annulus corresponds to a circle centered on the galaxy's center when the galaxy 
is viewed face-on.  An example of the final footprint of these annuli is shown 
in Figure \ref{fig:contour_plot_example}.

\begin{figure}
    \includegraphics[width=0.5\textwidth]{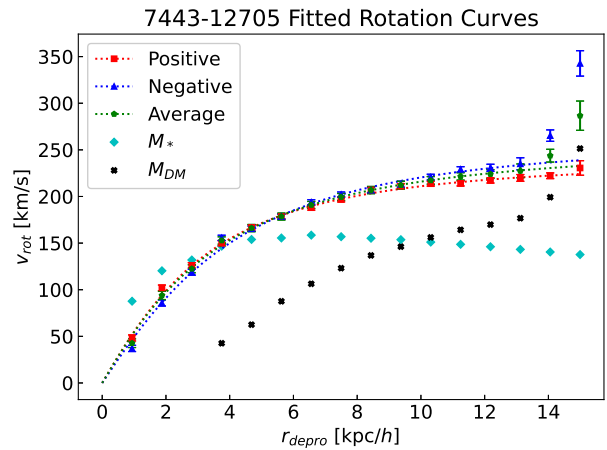}
    \caption{An example of the rotation curves for one galaxy.  The rotation 
    curve from the positive velocities is shown with red squares, the negative 
    velocities is shown with blue triangles, and the average velocity curve is 
    shown with green pentagons.  The best fits for these three curves are shown 
    in the corresponding colored dashed lines.  The rotation curve due to the 
    stellar mass component is shown in the teal diamonds.  The rotation curve 
    due to the estimated dark matter component is shown in the black x's.}
    \label{fig:rot_curve_example}
\end{figure}

From these velocities, we get a rotation curve for the side of the galaxy 
rotating toward us (the ``negative'' curve), and a rotation curve for the side 
rotating away from us (the ``positive'' curve).  We combine these two rotation 
curves by computing the average of the absolute value of the velocities at each 
deprojected radius to obtain a single rotation curve for the galaxy.  An example 
of these curves for one galaxy is shown in Figure \ref{fig:rot_curve_example}, 
with the rotation curve from the positive velocities drawn with red squares, the 
absolute value of the negative velocities drawn with blue triangles, and the 
average velocity curve drawn with green pentagons.  We fit each of these 
rotation curves with the parameterization defined by 
\cite{BarreraBallesteros18},
\begin{equation}\label{eqn:rot_curve}
    V_{\text{rot}} = \frac{V_{\text{max}} r_{\text{depro}}}{(R^\alpha_{\text{turn}} + r^\alpha_{\text{depro}})^{1/\alpha}}
\end{equation}
where $V_{\text{rot}}$ is the rotational velocity measured at a given 
deprojected radius $r_{\text{depro}}$, $V_{\text{max}}$ is the magnitude of the 
velocity on the plateau of the rotation curve, and $R_{\text{turn}}$ is the 
deprojected radius at which the curve plateaus.  The free parameters in the fit 
are $V_{\text{max}}$, $R_{\text{turn}}$ and $\alpha$.  It is important to 
remember that the MaNGA data only observe the extent of the luminous matter in 
each galaxy; as a result, our measured rotation curves can only extend to the 
edge of the stars in the galaxy.  

\begin{figure}
    \centering
    \includegraphics[width=0.5\textwidth]{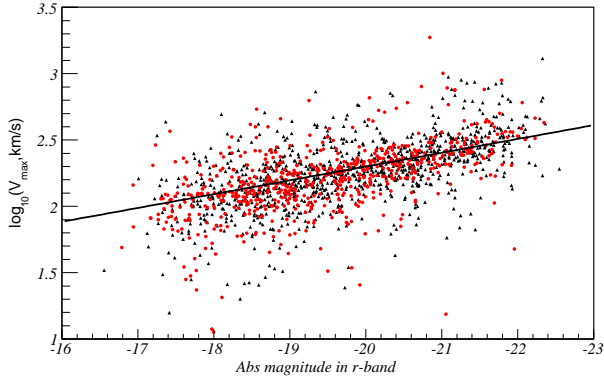}
    \caption{Logarithm of maximum velocity as a function of absolute magnitude 
    in the $r$-band for void (red circles) and wall (black triangles) galaxies.  
    The Tully-Fisher relation from \cite{Tully77} is shown in the solid black 
    line.}
    \label{fig:tully_fisher}
\end{figure}

To confirm that Equation \ref{eqn:rot_curve} successfully reproduces the form of 
the rotation curves for the galaxies in our sample, we plot the distribution of 
maximum velocities as a function of $M_{r}$ 
\citep[the Tully-Fisher relation;][]{Tully77}.  As seen in Figure 
\ref{fig:tully_fisher}, there is a strong positive correlation between the 
absolute magnitude and the maximum velocity, as expected.

\subsection{Best-fit Quality Cuts}\label{sec:curve_cuts}

We require that a galaxy's rotation curve contain at least four data points for 
the fitting procedure and analysis to proceed.  In addition, we only include 
those galaxies with the goodness of fit of the rotation curve characterized by a 
reduced $\chi^2$ ($\chi_\nu^2$) $< 10$.  Here, $\chi_\nu^2$ is normalized by the 
difference of the number of data points in the fit and the number of degrees of 
freedom of the fit.  If $\chi_\nu^2$ is too large for the average rotation 
curve, the fit to the positive curve is checked.  If it has a good fit 
($\chi_\nu^2 < 10$), then the positive curve's fit is used for the analysis; if 
not, then the negative curve is checked.  If none of these three fits have a 
$\chi_\nu^2 < 10$, then the last data point in each of the rotation curve data 
is removed, and the fits are re-calculated.  This procedure continues until 
either a good fit is found, or there are less than four data points remaining in 
the rotation curves, at which point the galaxy is removed from the sample.

Both this procedure and the removal of galaxies without smooth velocity maps 
(described in Section \ref{sec:masking}) preferentially eliminate galaxies with 
low dark matter content (for which a rotation curve is dominated by the bulge's 
mass) and elliptical galaxies (which lack organized rotational motion).  Because 
most elliptical galaxies are extremely bright, this introduces a selection bias 
toward fainter magnitudes.  Of the 4815 galaxies with IFU spectra available in 
the Pipe3D analysis of the SDSS DR15 MaNGA survey, we have rotation curves for 
\Nvoid void galaxies and \Nwall wall galaxies.  The number of galaxies with 
rotation curves classified as uncertain is \Nedge.

\subsection{Estimating the ratio of dark matter halo mass to stellar mass}\label{sec:mass_decomp}

The total mass internal to the galaxy is calculated using Equation 
\ref{eqn:M_within_r}, evaluated at the maximum deprojected radius measured in 
the rotation curves and using $V_{\text{max}}$ found from the best-fit rotation 
curve of Equation \ref{eqn:rot_curve}.

In order to compute the total dark matter halo mass, $M_{\text{DM}}$, and total 
stellar mass, $M_*$, within a galaxy, we need to decompose the galaxy's rotation 
curve into its constituent parts.  We estimate the total stellar mass contained 
within a particular radius by summing the stellar mass densities in each spaxel 
for all spaxels contained within that radius.  From this mass distribution, we 
can calculate the expected rotation curve due to the stellar mass component of 
the galaxy.  The rotation curve due to the stellar mass component is shown in 
the teal diamonds in Figure \ref{fig:rot_curve_example}.  By taking the stellar 
mass internal to the maximum deprojected radius, we can get an estimate of the 
discernible stellar mass in the galaxy.

Finally, the total dark matter halo mass for the galaxy is found by subtracting 
the total stellar mass estimated from the total mass calculated.  The rotation 
curve due to the dark matter component of the sample galaxy can be found in 
Figure \ref{fig:rot_curve_example} as the black x's. 

From this estimate of the total stellar mass and total dark matter halo mass 
observed for a given galaxy with this data set, we calculate the ratio of dark 
matter halo mass to stellar mass for each galaxy.

\subsection{Distribution of \Mratio}

\begin{figure}
    \centering
    \includegraphics[width=0.5\textwidth]{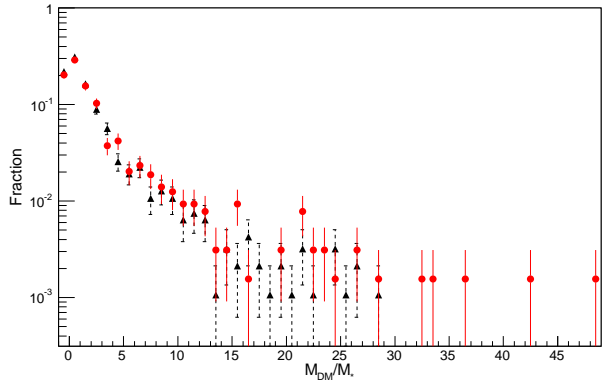}    
    \caption{Distribution of the ratio of dark matter halo mass to stellar mass 
    for void (red circles, solid line) and wall (black triangles, dotted line) 
    galaxies.  There is no statistically significant difference in the 
    distributions between the two environments.}
    \label{fig:dm_sm_hist}
\end{figure}

The distributions of the ratio of dark matter halo mass to stellar mass, 
\Mratio, for void and wall galaxies used in this analysis are shown in Figure 
\ref{fig:dm_sm_hist}.  There is no statistically significant difference in the 
distributions of \Mratio between the two environments, similar to the results of 
\cite{Duckworth19}.  However, these results contradict the simulation 
predictions of \cite{Martizzi19} and theoretical predictions of 
\cite{Arkhipova07}, who both found that void galaxies should contain larger 
fractions of dark matter than those in denser regions (filaments and knots).

\begin{figure*}
    \centering
    \includegraphics[width=0.9\textwidth]{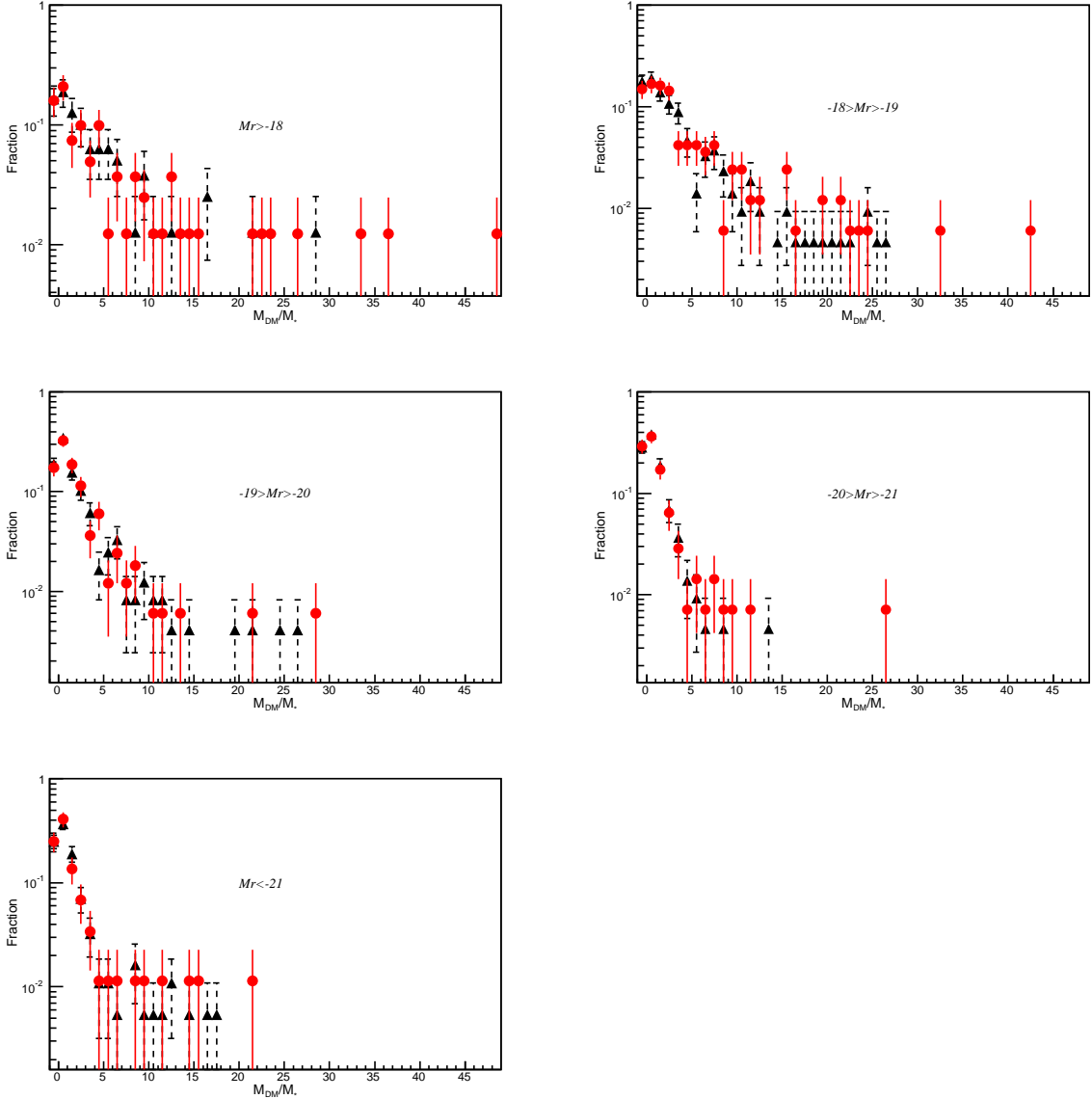}
    \caption{Distributions of \Mratio for void (red circles, solid line) and 
    wall (black triangles, dotted line) galaxies, binned by absolute magnitude.  
    There is no statistically significant difference between the environments in 
    any of the magnitude ranges.}
    \label{fig:dm_sm_hist_Mr}
\end{figure*}

We acknowledge that \Mratio has the potential to vary as a function of galaxy 
size, so we also look at the distributions when the galaxies are binned by 
absolute magnitude.  Shown in Figure \ref{fig:dm_sm_hist_Mr} and quantified in 
Table \ref{tab:shifts}, the void environment does not affect the mass ratio 
distribution in any galaxy magnitude range.  In Table \ref{tab:shifts}, we 
present the sample size of the galaxies binned by their absolute magnitude and 
the probability of $\chi^2$ of the difference in the distributions over \Mratio 
between the void and wall galaxies.  The lowest probability of $\chi^2$ 
corresponds to the inclusive sample, where the median ratio of the dark matter 
halo mass to stellar mass is $1.03\pm 0.08$ in voids and $0.89\pm 0.07$ in 
walls.  Once galaxies are binned by their absolute magnitude, their 
distributions in \Mratio are even more consistent between the two environments.  
These results do not support those of \cite{Tojeiro17}, who found that the ratio 
of halo mass to stellar mass is larger for small galaxy groups in voids than 
those in more dense environments.

\begin{deluxetable}{cccc}
    \tablewidth{0pt}
    \tablecolumns{4}
    \tablecaption{Sample sizes and the probability of $\chi^2$ of the agreement 
    between the distributions in the ratio of dark matter halo mass to stellar 
    mass of the void and wall galaxies, binned by absolute magnitude as indicated 
    on the left.\label{tab:shifts}}
    \tablehead{\colhead{Range in $M_r$} & \colhead{Void} & \colhead{Wall} & \colhead{$P(\chi^2)$}}
    \startdata
        All                     & \Nvoid & \Nwall & 0.88 \\
        Fainter than $-18$      & 81     & 79     & 0.99 \\
        Between $-18$ and $-19$ & 167    & 215    & 0.97 \\
        Between $-19$ and $-20$ & 166    & 243    & 0.99 \\
        Between $-20$ and $-21$ & 140    & 217    & 0.99 \\
        Brighter than $-21$     & 88     & 184    & 0.99 \\
    \enddata
\end{deluxetable}

\subsection{Relationship between \Mratio and $M_r$}

\begin{figure}
    \centering
    \includegraphics[width=0.5\textwidth]{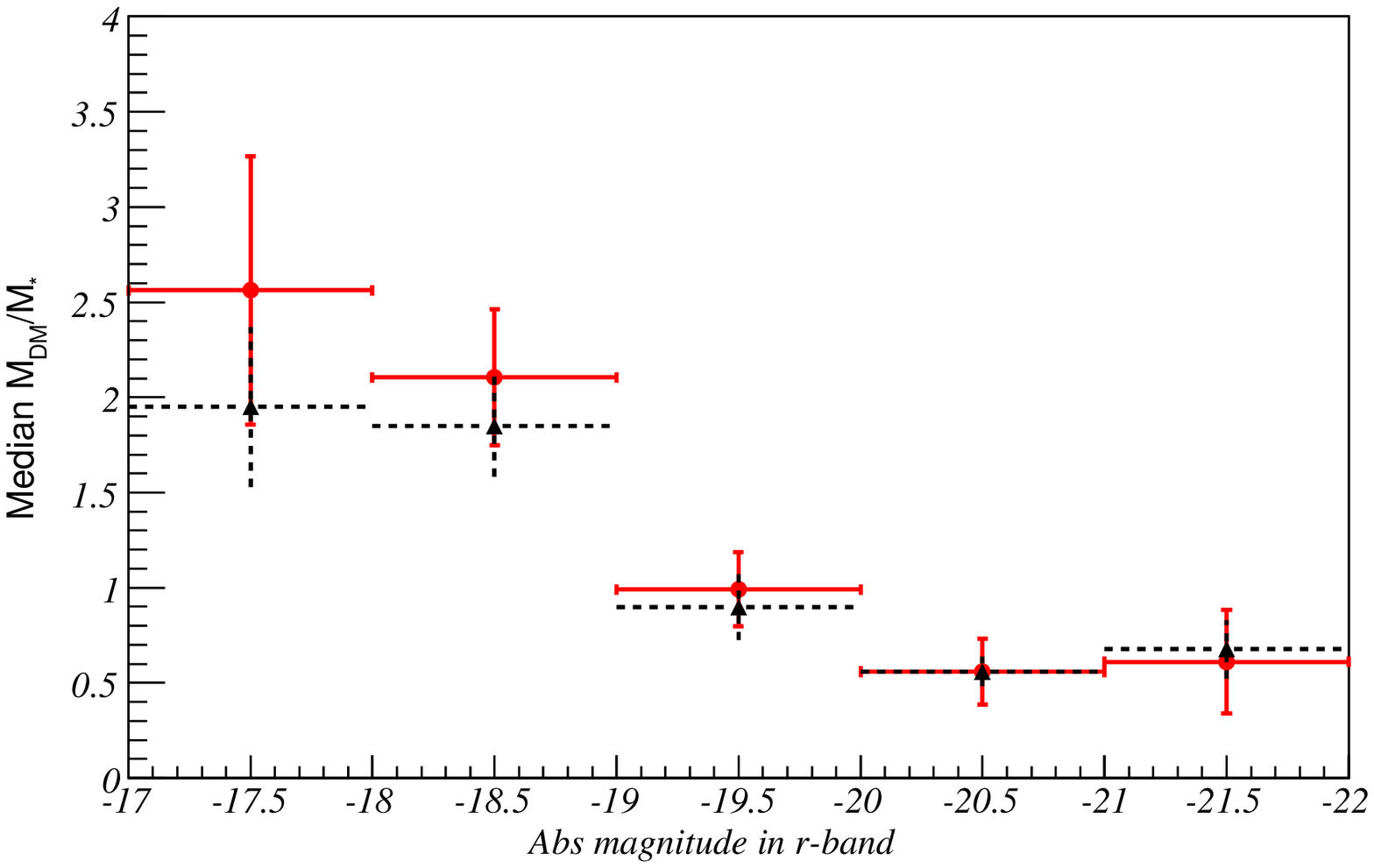}
    \caption{Median dark matter ratio as a function of absolute magnitude for 
    void (red circles, solid line) and wall (black triangles, dotted line) 
    galaxies.  The fraction of dark matter decreases with increasing brightness 
    in all but the brightest bins, independent of environment.}
    \label{fig:MedianDMratio_Mr}
\end{figure}

The relationship between \Mratio and $M_r$ (a form of the stellar-halo mass 
relation, SHMR) is shown in Figure \ref{fig:MedianDMratio_Mr}.  The 
distributions seen in Figure \ref{fig:dm_sm_hist_Mr} show that fainter galaxies 
contain higher fractions of dark matter relative to their stellar mass content 
than brighter galaxies.  This trend is shown more clearly in Figure 
\ref{fig:MedianDMratio_Mr}, where we investigate the relationship between the 
median mass ratio in each magnitude bin.  Fainter galaxies (those with 
$M_r > -18$) have the highest \Mratio, and the mass ratio decreases with the 
galaxies' magnitudes.  A similar relationship is seen in \cite{Persic96}.  These 
distributions show no statistically significant dependence on the environment.

In the brightest galaxies, we see that \Mratio does not follow the trend present 
in the other magnitude bins.  Instead, the mass ratio appears to either remain 
constant or increase slightly, indicating that the brightest galaxies 
($M_r < -21$) have more dark matter halo mass relative to their stellar mass 
than slightly fainter galaxies.  Similar trends are seen in \cite{Moster10} and 
\cite{Paolo18}, where the SHMR is found to peak around 
$\log(M_*/M_\odot) \sim 10.5$.

\begin{figure}
    \centering
    \includegraphics[width=0.5\textwidth]{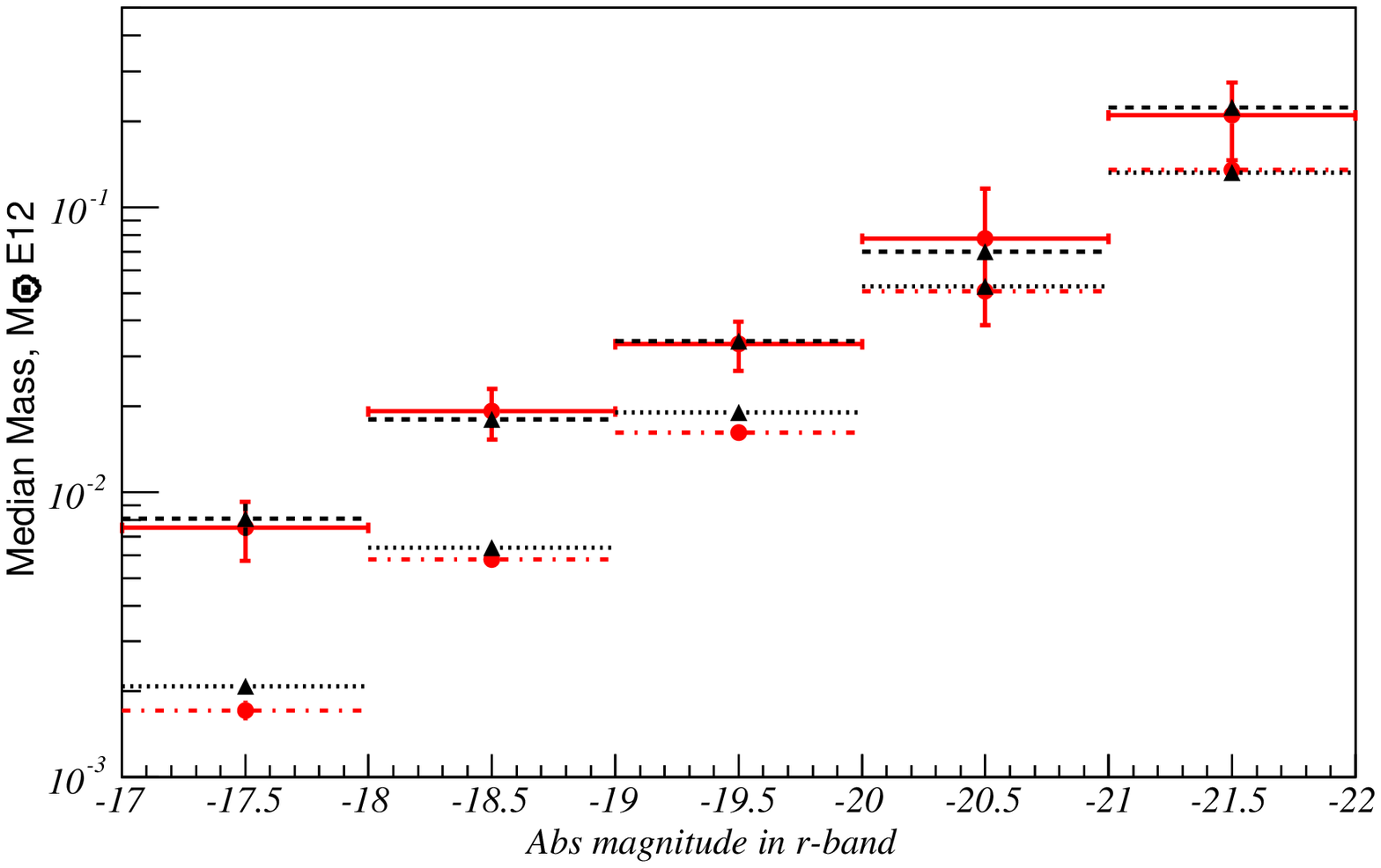}
    \caption{Median total mass (solid/dotted lines) and stellar mass 
    (dashed/dot-dashed lines) as functions of absolute magnitude for void (red 
    circles) and wall (black triangles) galaxies.  Both the total mass and 
    stellar mass increase with brightness, as expected.}
    \label{fig:Mtot_Mstar_vs_Mr}
\end{figure}

The dependencies of the median total mass and median stellar mass on the 
absolute magnitude are presented in Figure \ref{fig:Mtot_Mstar_vs_Mr}.  As 
expected, both the total and stellar mass increase with galaxy brightness.  The 
difference between the total mass and stellar mass in a magnitude bin 
corresponds to the dark matter halo mass.  Figure \ref{fig:Mtot_Mstar_vs_Mr} 
shows that this difference increases with decreasing brightness in all but the 
brightest galaxy bin, and the faintest galaxies have the largest difference 
between their total mass and stellar mass.  This matches the relationship we see 
in Figure \ref{fig:MedianDMratio_Mr}, where the ratio of dark matter halo mass 
to stellar mass decreases with increasing brightness in all but the brightest 
magnitude bin.

\begin{figure}
    \centering
    \includegraphics[width=0.5\textwidth]{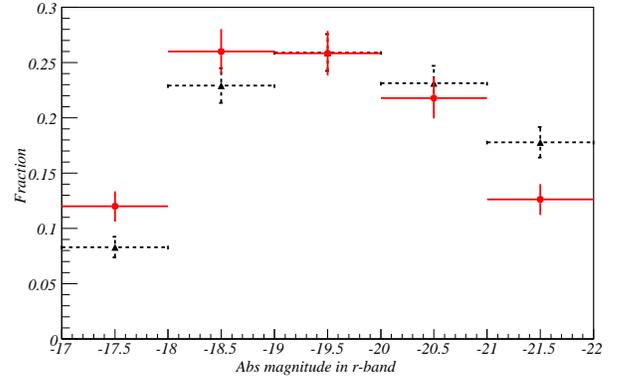}
    \caption{Distribution of void (red circles, solid line) and wall (black 
    triangles, dotted line) galaxies in absolute magnitude.  Void galaxies are 
    characteristically fainter than galaxies found in denser regions.}
    \label{fig:Fraction_mag}
\end{figure}

It is well known \citep{Hoyle05} that void galaxies are characteristically 
fainter than wall galaxies.  To confirm that our sample follows this trend, we 
show the normalized distributions in absolute magnitude in the two environments 
in Figure \ref{fig:Fraction_mag}.  Indeed, voids contain a higher fraction of 
faint galaxies than denser regions.  Therefore, if \Mratio depends on a galaxy's 
brightness, then the lower probability of $\chi^2$ in the inclusive bin (Figure 
\ref{fig:dm_sm_hist}, first row of Table \ref{tab:shifts}) is explained by the 
different distributions in the brightness of void and wall galaxies.  In 
general, there is a higher fraction of dwarf galaxies with higher dark matter 
content in the voids, while walls contain brighter galaxies with lower dark 
matter content.

\section{The influence of the void environment}
\subsection{Dark matter content of fainter galaxies}

The well-known SHMR predicts dwarf galaxies to have a higher fraction of dark 
matter than more massive galaxies \citep[and references therein]{Wechsler18}.  
Previous observations of the dwarf galaxies in the Local Group \citep{Mateo98} 
indicate that there is an inverse correlation between the luminosity and 
mass-to-light ratio in dwarf spheroidal galaxies.  In addition, 
\cite{TorresFlores11} found that low-mass galaxies in the GHASP survey are 
dominated by dark matter, while those with higher masses are dominated by 
baryonic matter.  Recent simulations by \cite{Martin19} showed that 
low-surface-brightness galaxies have a slightly higher fraction of dark matter 
compared to high-surface-brightness galaxies.  Our results in Figure 
\ref{fig:MedianDMratio_Mr} align with the SHMR, where the relative fraction of 
dark matter increases with decreasing luminosity.

The deviation from a constant SHMR is suggested to be the result of various 
feedback processes that reduce the star formation efficiency: supernovae 
feedback in lower-mass galaxies, and AGN feedback in the most massive galaxies 
\citep{Paolo18,Wechsler18}.  We see no statistically significant effect from the 
void environment on the relationship between the median ratio of dark matter 
halo mass to stellar mass and the absolute magnitude, so the source of this 
relationship should not be influenced by the large-scale environment.

$\Lambda$CDM cosmology predicts that star formation commenced at a later time in 
void galaxies than for galaxies in denser regions 
\citep{Gottlober03,Goldberg05,Cen11}.  If this is true, then the SHMR for void 
galaxies should be similar to that for galaxies in denser regions at an earlier 
redshift.  The lack of a statistically significant difference in the 
relationships between the ratio of dark matter halo mass to stellar mass and the 
absolute magnitudes of galaxies in voids and in denser regions shown in Figure 
\ref{fig:MedianDMratio_Mr} aligns with the simulation results of 
\cite{Behroozi19}, who showed that there is not much variance in the SHMR at 
redshifts $z \lesssim 4$.

\subsection{Implications for the gas-phase metallicity}

Studies by \cite{Douglass17b} and \cite{Douglass18} suggest that a shift toward 
higher \Mratio in void galaxies could explain why void dwarf galaxies have 
gas-phase metallicities similar to those of dwarf galaxies in denser regions.  
However, as seen in Figures \ref{fig:dm_sm_hist} and \ref{fig:dm_sm_hist_Mr} and 
in Table \ref{tab:shifts}, there is no statistically significant difference in 
the mass ratios for galaxies as a function of their large-scale environment.

\begin{figure}
    \centering
    \includegraphics[width=0.5\textwidth]{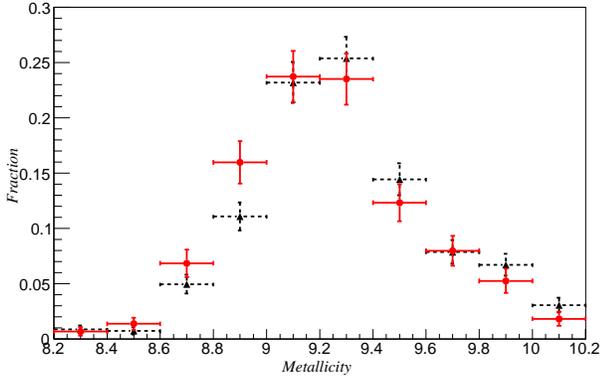}
    \caption{Gas-phase metallicity (\logOH) distribution for void (red circles, 
    solid line) and wall (black triangles, dotted line) galaxies.  The apparent 
    shift toward lower metallicities in void galaxies is a result of the larger 
    fraction of fainter galaxies in voids and the mass-metallicity relation.}
    \label{fig:Metallicity}
\end{figure}

\begin{figure}
    \centering
    \includegraphics[width=0.5\textwidth]{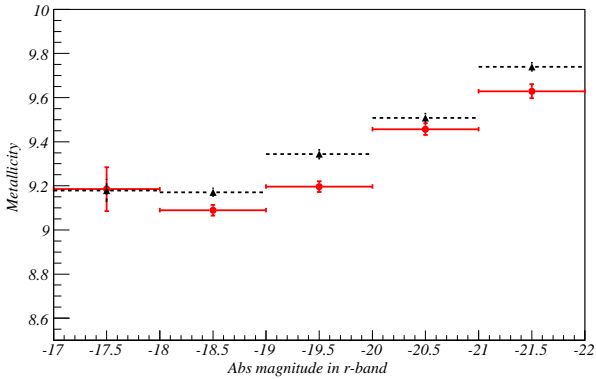}
    \caption{Average gas-phase metallicity (\logOH) for galaxies binned by $M_r$ 
    for void (red circles, solid line) and wall (black triangles, dashed line) 
    galaxies.  The typical mass-metallicity relationship (increasing metallicity 
    with absolute magnitude) is observed.}
    \label{fig:MET_vs_Mr}
\end{figure}

Using the N2O2 diagnostic as calibrated by \cite{Brown16}, we calculate the 
gas-phase metallicities (\logOH) of the galaxies in our sample using emission 
line flux values from the MPA-JHU value-added catalog,\footnote{Available at 
\url{http://www.mpa-garching.mpg.de/SDSS/DR7/}} which is based on the SDSS DR7 
sample of galaxies.  The distribution over metallicity is shown in Figure 
\ref{fig:Metallicity}.  The overall shift toward lower metallicities in the void 
galaxies is a result of the larger fraction of faint galaxies present in voids 
and the known mass-metallicity relation, as shown in Figure \ref{fig:MET_vs_Mr}.

\begin{figure}
    \centering
    \includegraphics[width=0.5\textwidth]{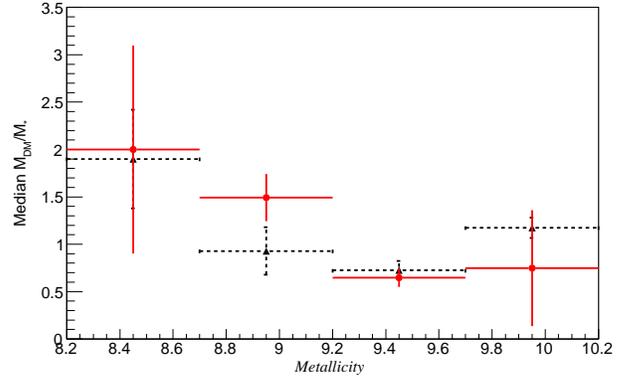}
    \caption{Ratio of dark matter halo mass to stellar mass as a function of 
    gas-phase metallicity (\logOH) for void (red circles, solid line) and wall 
    (black triangles, dotted line) galaxies.  As with the SHMR, \Mratio 
    decreases with increasing metallicity in all but the highest metallicity 
    bin.}
    \label{fig:DM_vs_Met}
\end{figure}

We also study the dependence of \Mratio on metallicity, shown in Figure 
\ref{fig:DM_vs_Met}.  Due to the mass-metallicity relation (Figure 
\ref{fig:MET_vs_Mr}) and the observed relationship between a galaxy's absolute 
magnitude and its dark matter fraction (Figure \ref{fig:MedianDMratio_Mr}), we 
expect that galaxies with lower metallicities will have higher mass ratios.  
Indeed, this is what is shown in Figure \ref{fig:DM_vs_Met}, mimicking the 
relationship between $M_r$ and \Mratio seen in Figure 
\ref{fig:MedianDMratio_Mr}.  There is no statistically significant difference in 
the relationships between the gas-phase metallicity and \Mratio of void galaxies 
and galaxies in denser regions.  In general, galaxies with lower metallicities 
tend to have a higher fraction of dark matter.

\subsection{Shape of the dark matter halo profile}

\begin{figure}
    \centering
    \includegraphics[width=0.5\textwidth]{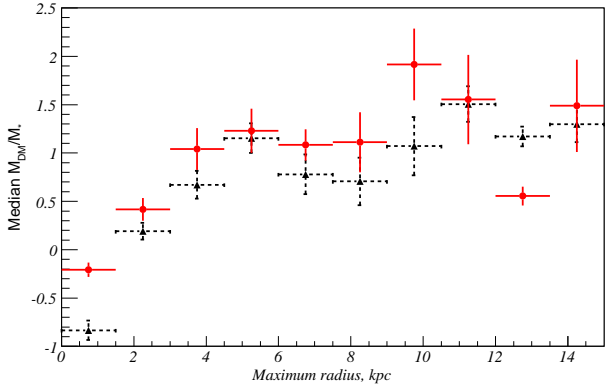}
    \caption{Ratio of dark matter halo mass to stellar mass as a function of 
    maximum radius obtained from the fit of the rotation curves for void (red 
    circles, solid line) and  wall (black triangles, dotted line) galaxies.  
    Both void and wall galaxies continue to exhibit similar \Mratio values as we 
    probe closer to the edge of the galaxy.}
    \label{fig:DM_vs_radius}
\end{figure}

If the dark matter halo profile is the same for void and wall galaxies, then any 
difference in the dark matter halo mass would be accounted for beyond the extent 
of the stellar disk.  SDSS MaNGA only observes the extent of the luminous matter 
in the galaxies, so we are not able to detect any differences beyond the extent 
of the stars.  The dark matter halo mass estimates we calculate from the 
rotation curves provide a lower limit on the dark matter halo mass for these 
galaxies.  As a result, we would not see any difference in the ratio of dark 
matter halo mass to stellar mass in the MaNGA galaxies if the shape of the dark 
matter halo is the same.

If instead, the dark matter halo is more shallow in void galaxies (more 
pancake-shaped) than in wall galaxies, we would expect to see lower ratios of 
dark matter halo mass to stellar mass in the void dwarf galaxies with these 
data based on the conclusions of \cite{Douglass17b} and \cite{Douglass18}.  Our 
results eliminate this potential geometry of the dark matter halo.  If there is 
a difference in the halo profile of void galaxies, then it is in the extent of 
the halo beyond the luminous edge of the galaxy.  We show the relationship 
between the ratio of dark matter halo mass to stellar mass and the maximum 
radius probed by the fit rotation curves in Figure \ref{fig:DM_vs_radius}.  Both 
void and wall galaxies continue to exhibit similar \Mratio values with 
increasing $R_{\text{max}}$.  If there is any difference in the dark matter halo 
masses between void and wall galaxies with similar stellar masses, it is far 
beyond the edge of the luminous matter.

\section{Conclusions}

We investigate the influence of the void environment on the fraction of dark 
matter in galaxies by comparing the ratio of dark matter halo mass to stellar 
mass of galaxies in voids with galaxies in denser regions.  Using the H$\alpha$ 
velocity maps from the SDSS MaNGA DR15, we are able to measure the rotation 
curves of \Nvoid void galaxies and \Nwall galaxies in denser regions.  From the 
rotation curves, we estimate the total mass of the galaxies; combined with the 
galaxies' stellar masses, we infer their dark matter halo masses.

We find that there is no difference in the ratio of dark matter halo mass to 
stellar mass in void galaxies and in galaxies in denser regions.  We also do 
not find a statistically significant difference in either the relationship 
between the mass ratio and the absolute magnitude or in the relationship between 
the gas-phase metallicity (\logOH) and the mass ratio.

When separated by absolute magnitude, we find a general trend toward a 
decreasing ratio of dark matter halo mass to stellar mass with increasing 
absolute magnitude in all but the brightest galaxies.  This relationship is 
independent of a galaxy's large-scale environment, so the SHMR does not appear 
to be affected by the void environment.

Because the MaNGA IFU only cover the extent of the luminous component of the 
galaxies, the dark matter halo masses that we estimate are lower limits on the 
total dark matter halo masses.  Therefore, our results indicate that the dark 
matter halo profile is similar in void galaxies and in galaxies in denser 
regions.  We cannot make any conclusions about the extent of the dark matter 
halo beyond the limit of the luminous matter.  Further analysis with \ion{H}{1} 
data, for example, would help to discern any environmental affect on the extent 
of the dark matter halo.

\section{Acknowledgements}

The authors would like to thank both Michael Vogeley for his insightful comments 
and Stephen W. O'Neill, Jr. for his help in determining the smoothness of the 
velocity maps.  J.A.S. and R.D. acknowledge support from the Department of 
Energy under the grant DE-SC0008475.0.

This project makes use of the MaNGA-Pipe3D data products.  We thank the IA-UNAM 
MaNGA team for creating this catalog, and the Conacyt Project CB-285080 for 
supporting them.

Funding for the Sloan Digital Sky Survey IV has been provided by the Alfred P. 
Sloan Foundation, the U.S. Department of Energy Office of Science, and the 
Participating Institutions.  SDSS-IV acknowledges support and resources from the 
Center for High-Performance Computing at the University of Utah.  The SDSS web 
site is www.sdss.org.

SDSS-IV is managed by the Astrophysical Research Consortium for the 
Participating Institutions of the SDSS Collaboration including the Brazilian 
Participation Group, the Carnegie Institution for Science, Carnegie Mellon 
University, the Chilean Participation Group, the French Participation Group, 
Harvard-Smithsonian Center for Astrophysics, Instituto de Astrof\'isica de 
Canarias, The Johns Hopkins University, Kavli Institute for the Physics and 
Mathematics of the Universe (IPMU) / University of Tokyo, the Korean 
Participation Group, Lawrence Berkeley National Laboratory, Leibniz Institut 
f\"ur Astrophysik Potsdam (AIP),  Max-Planck-Institut f\"ur Astronomie (MPIA 
Heidelberg), Max-Planck-Institut f\"ur Astrophysik (MPA Garching), 
Max-Planck-Institut f\"ur Extraterrestrische Physik (MPE), National Astronomical 
Observatories of China, New Mexico State University, New York University, 
University of Notre Dame, Observat\'ario Nacional / MCTI, The Ohio State 
University, Pennsylvania State University, Shanghai Astronomical Observatory, 
United Kingdom Participation Group, Universidad Nacional Aut\'onoma de M\'exico, 
University of Arizona, University of Colorado Boulder, University of Oxford, 
University of Portsmouth, University of Utah, University of Virginia, University 
of Washington, University of Wisconsin, Vanderbilt University, and Yale 
University.


\bibliographystyle{aasjournal}
\bibliography{Doug0620_sources}

\end{document}